# Hardening-softening transition in Fe-doped Pb(Zr,Ti)O$_3$ ceramics and evolution of the third harmonic of the polarization response

Maxim I. Morozov

*Institute of Materials Science, Dresden University of Technology, D-01062 Dresden, Germany*

Dragan Damjanovic[a]

*Ceramics Laboratory, Swiss Federal Institute of Technology - EPFL,*

*CH-1015 Lausanne, Switzerland*

**Abstract.** It is shown that the phase angle of the third harmonic of polarization response ($\delta_3$) is very sensitive to the ageing state of hard, Fe-doped lead zirconate titanate [Pb(Zr,Ti)O$_3$ or PZT] ceramics and may thus provide rich information on processes responsible for ageing in hard ferroelectrics. When hard PZT ceramics experience hardening (aged state) - softening (deaged state) transition, $\delta_3$ changes by about 90° at subswitching fields and by about 180° at switching conditions. Evolution of $\delta_3$ with time, temperature and electric field amplitude during deageing suggests that at least two mechanisms of charge migration, one short-range and one long-range, may participate in the deageing process.

---

[a] Electronic mail: dragan.damjanovic @epfl.ch

1. **Introduction.**

Ageing is one of the key properties that distinguish between hard (acceptor doped) and soft (donor doped) ferroelectric materials. It is well known from the early works in piezoelectric ceramics that properties of hard materials decrease with time considerably faster than properties of soft materials[1]. Understanding origin of ageing in hard and its absence in soft materials may thus help identifying details of hardening and softening mechanisms. Electric properties of hard ferroelectrics change reversibly, i.e. samples can be deaged by heating up over the ferroelectric-paraelectric phase transition temperature[2] or by applying cyclic electric field[3]. When the external influence is removed, the sample ages again. The mechanisms of ageing are usually related to presence of mobile charged species such as point defects or defect complexes, which stabilize the domain pattern and decrease the domain wall contribution to the polarization response. The effect of mobile charged species in hard PZT is evident from many experimental studies including observation of defect states by the electron paramagnetic resonance (EPR)[4-6], analysis of temperature dependence of polarization (P) - electric field (E) loops[3] (also combined with a TEM study[7]), and the dielectric spectroscopy.[8] However, the exact mechanism dominating the stabilization of the domain walls is still disputable.[2,3,7,9-15] There are three main models usually considered:[3] (i) the bulk effect (caused by alignment of charged defects with polarization within ferroelectric domains);[3,15] (ii) the domain wall effect (diffusion of charges towards domain walls thus creating pinning centers);[16-18] and (iii) surface effect (drift and gathering of charges near grain boundaries and other interfaces). The bulk and domain walls models are based on results of experimental studies[3-6,9,11-16] while the dominating contribution of surface (interface) effect was proposed by modeling in Refs. 10,19 and in Ref. 3.

The microscopic processes responsible for the ageing are most clearly manifested macroscopically in pinching (or constriction) of *P-E* hysteresis loops at zero field, as shown in Fig. 1. During ageing the loop becomes increasingly more pinched while depinching (relaxation) of the loop takes places during deageing, Fig. 1. In deaged state a hard material appears macroscopically as soft (pinching disappears and ferroelectric hysteresis loop exhibits a "square" shape).[2,3] In their seminal paper[3], Carl and Härdtl have shown that deageing can be put in evidence by monitoring the separation of the switching current peaks during the field cycling. A well aged sample is characterized by a pinched hysteresis loop and two pairs of peaks in the switching current (one pair with positive, the other with negative peaks), each pair corresponding to two distinct parts of the loop. As deageing progresses, the positions of peaks shift and the two positive and two negative peaks merge. When the pinching disappears only two peaks are left, the loop is said to be "relaxed" and the sample considered to be deaged.

. In this study we investigate the ageing-deageing processes in hard lead zirconate titanate [Pb(Zr,Ti)O$_3$ or PZT] ceramics using the harmonic analysis of the polarization response.[20] It is shown that the phase angle of the third harmonic of polarization response is very sensitive to the ageing state of the sample. Since properties of ferroelectric materials are naturally described using nonlinear theories this approach offers a clear, experimentally verifiable criterion for testing models of ageing and deageing processes and thus theories of hardening mechanisms. Representation of an *ac*-signal in the form of harmonic amplitudes and phase angles is also convenient for monitoring the evolution of nonlinear response as a function of different variables, such as time, temperature, and driving field amplitude. Finally, if the apparent nonlinearity of material response is a result of several contributions differently depending on time, the harmonic analysis in frequency domain may be very helpful in separating such phenomena. For example, nonlinearity imposed by ageing through alignment of dipoles or migration of charges may be superimposed on and have different time dependences than other nonlinear contributions such as intrinsic response and domain-walls dynamics. Harmonic analysis combined with the phase data for each harmonic has an additional advantage of giving directly information on the hysteretic behavior of the system. Anhysteretic part (e.g., intrinsic lattice response) results in in-phase response with respect to the driving field. The hysteretic (out-of-phase) component may be due to the nonlinear behavior,[21,22] dielectric losses, conductivity of the sample, polarization of interfaces and inductive wires in the circuit. At electric field amplitudes used in this work it is assumed that nonlinearities of the lattice are small in comparison with those related to domain walls displacement (which is in turn controlled by defect dipoles reorientation).[15,23-26] Indeed, doping PZT with an atomic percent of acceptor dopant or less would not be expected to affect lattice properties so much as to cause drastic changes in the hysteresis loops; it is reasonable to invoke domain walls related effects.

Properties of PZT ceramics are examined at "subswitching" and "switching" *ac* fields. These fields are defined roughly with respect to the coercive field that would be measured in a completely polarized sample. An illustration is given in Fig. 1 of Ref. 27.

"Switching fields" are higher than the coercive field of the sample. At switching fields the remanent polarization of the fully polarized sample, $P_R$, will switch from $P_R$ to $-P_R$ or will be substantially modified. "Subswitching" fields are lower than the coercive field of the sample. At subswitching fields the remanent polarization of the sample is not substantially modified although individual domain walls may switch. These fields depend on temperature, driving field frequency, time, and sample details (e.g., grain size, degree of ageing).

No specific model of ageing is assumed in this paper. The ensuing discussion of experimental results is, however, based on idea of rearrangement of defect charges. Term "dipoles" is used for convenience and may refer to both dipoles formed between near-neighbor defect centers of opposite charges or features associated with longer-range separation of charges.

**2. Samples preparation**

The samples of hard rhombohedral $Pb(Zr_{0.58}Ti_{0.42})O_3$ ceramics doped with $Fe^{+3}$ were prepared by conventional solid state process using standard mixed oxide route. $Fe_2O_3$ powder was mixed with stoichiometric amounts of PbO, $TiO_2$ and $ZrO_2$ precursors (compensated for water content, mostly in $ZrO_2$). The dopant substitution is assumed on (Zr,Ti) site so that the nominal formula of the samples is $Pb(Zr_{0.58}Ti_{0.42})_{1-x}Fe_xO_3$ with x = 0.1, 0.5 and 1.0 % (abbreviated 58/42 PZT with x at% Fe). Powders were calcined in lead oxide saturated alumina crucibles covered by alumina plates. Powders were milled and sieved before and after calcinations. The sintering was performed on pellets uniaxially pressed at 40 MPa and packed into covered alumina crucibles whose volume was a little larger than the volume of pellets. The inner space of the crucibles was filled up with the powder of the same composition as the pressed pellet in order to prevent intensive lead oxide evaporation during the thermal treatment. Ageing of samples was assured by their slow cooling within the furnace down to room temperature, at which samples were held for several days. Thermal quenching was performed by very fast cooling achieved by dropping samples into water from temperature of about 480 °C, which is well above the Curie temperature ($T_C \sim 360°C$). Gold electrodes were sputtered on two parallel faces of the samples.

## 3. Harmonic analysis

A sinusoidal voltage signal generating electric field $E = E_0 \sin(\omega t)$ was applied to one electroded face of the sample. The polarization charge from the opposite face was either measured directly using a charge amplifier or derived from the measured current through a resistor connected in series to the sample[8]. The harmonic analysis of the voltage signal from the charge amplifier or the voltage drop across the resistor was performed using a lock-in amplifier. It should be noted that the lock-in amplifier used (Stanford Research SR830 DSP) defines the input signal as $V_{sig} \sin(\omega t + \theta_{sig})$ where $\theta_{sig}$ refers to the phase angle between the driving and the measured signal. The phase angles shown are adjusted for the phase shift introduced by the charge amplifier (180°) or resistor (90°).

The harmonic amplitudes and phase angles of up to first nine harmonics of the polarization $P$ were characterized at both subswitching and switching amplitudes of electric field. The perfectly symmetrical response of material, where $P(E) = -P(-E)$, implies so-called half-wave symmetry[28] $P(\omega t) = -P(\omega t + \pi)$. Half-wave symmetry of polarization response is obtained in unpoled and non-textured ceramics with random orientation of grains. In that case the amplitudes of all even harmonics are ideally equal to zero. The polarization response can be expressed as:

$$P(t) = P_1' \sin(\omega t) + P_1'' \cos(\omega t) + P_3' \sin(3\omega t) + P_3'' \cos(3\omega t) + P_5' \sin(5\omega t) + P_5'' \cos(5\omega t) + ...$$
$$= \sum_{n=1,3,...}^{\infty} \sqrt{(P_n')^2 + (P_n'')^2} \sin[n\omega t + \delta_n] \qquad (1)$$

where $P_n'$ and $P_n''$ are the in-phase and out-of-phase amplitudes of polarization and $\delta_n = \arctan(P_n''/P_n')$ is the phase angle of the $n$th harmonic.

The electric field reaches the extrema $\pm E_0$ when $\omega t = \pm \pi/2$, and crosses zero when $\omega t = 0$ or $\pi$. At the same time the polarization response (1) is determined by either only in-phase or only out-of-phase components, as shown in Table I. The $P_n'$ and $P_n''$ as well as their ratios (phase angles), contain full information on the hysteresis loop geometry. In practice, even harmonics are present and may be non-negligible in poled samples.[29,30] This is not the case for unpoled samples where polarization response should ideally be dominated by odd harmonics. Application of the driving field may, however, polarize samples and induce even harmonics.[31] Under experimental conditions used in this work the amplitudes of even harmonics were small in comparison with that of odd harmonics and even harmonics were not considered.

Table I: Polarization in a nonlinear material with a half-wave symmetry at different peak values of the electric field

| Angle | Contribution to polarization |
|---|---|
| $\omega t = \pi/2$ | $P = +P'_1 - P'_3 + P'_5 - P'_7 + ...$ |
| $\omega t = \pi$ | $P = -P''_1 - P''_3 - P''_5 - P''_7 + ...$ |
| $\omega t = 3\pi/2$ | $P = -P'_1 + P'_3 - P'_5 + P'_7 + ...$ |
| $\omega t = 2\pi$ | $P = +P''_1 + P''_3 + P''_5 + P''_7 + ...$ |

**4. Results and discussions**

**4.1. Characterization at switching fields**

The experimental evidence of hysteresis loop relaxation is shown in Fig.1. At room temperature, a strongly hard (1 at% Fe) ceramic demonstrates very slow field-induced hysteresis loop relaxation; one is observed after long cycling time (more than 10 hours) at relatively high frequency (1 kHz) and high amplitude (about 25 kV/cm) of driving electric field (Fig.1a). For comparison, in experiments of Carl and Härdtl[3] a complete relaxation of Mn-doped rhombohedral PZT samples was observed at room temperature after 200 cycles with the field amplitude of 50 kV/cm and frequency 0.15 Hz. At higher temperatures the rate of hysteresis loop releasing is faster due to the thermally activating nature of the phenomenon.[3] As shown in Fig.1b, the measurements of *P-E* loop performed at 10 Hz reveal a non-constricted hysteresis loop at higher temperatures already at the first minute of cycling. Further opening of hysteresis loop towards the rectangular geometry typical for soft materials is observed with cycling at 125 °C and 10 Hz (Fig.1c). We point out that there are at least two reasons that limit the hysteresis loop opening at room temperature and high frequency cycling (Fig.1a). The first is insufficient energy for removing all factors contributing to the hysteresis loop pinching and slow kinetics of defects responsible for pinching. The second is ferroelectric fatigue[32-34] during cycling which leads to the loop suppression.

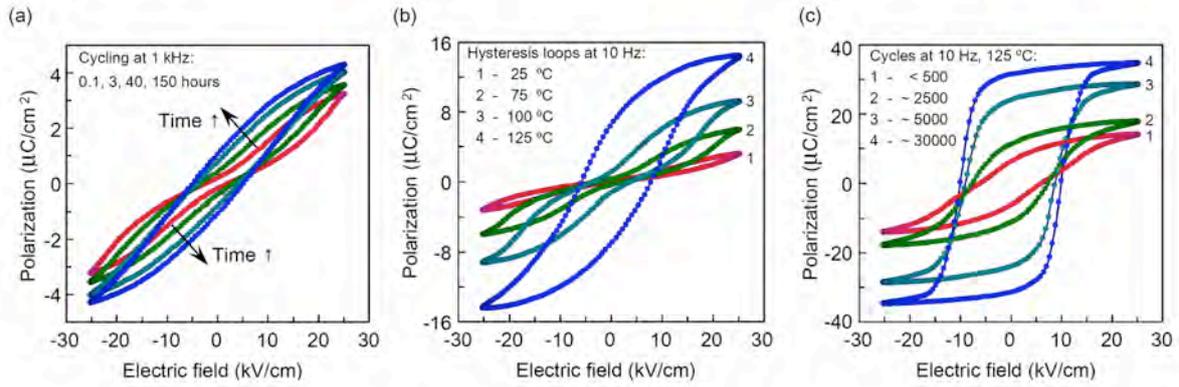

*Figure 1.* *Ferroelectric hysteresis loop relaxation observed in hard 58/42 PZT ceramics doped with 1.0 at.% Fe: (a) electric field cycling at high frequency (1kHz); (b) the thermal activation of hysteresis relaxation (at 10 Hz); (c) low frequency field cycling (10 Hz) at 125 °C.*

The deageing process in hard PZT ceramics under electric field cycling and the thermally activated nature of ageing were shown in Ref. 3 by analysis of time that is necessary for apparent disappearance of the constriction of the hysteresis loop at various temperatures. As mentioned in Introduction, the main criterion for determining this time in Ref. 3 was the moment when the switching current peaks merge. This method, however, is not always convenient because the current peaks may not be well pronounced. Especially in the peak merging region the peak separation may be poorly defined if ageing is weak. In addition, the information on nonlinear processes behind the deageing process is hidden within the current curves, which comprise contribution of all processes contributing to switching and ageing. In the following, an alternative approach for analysis of the deageing process using advantages of harmonic analysis is demonstrated.

The time evolution of the phase angles of the first nine harmonics of the polarization is shown in Fig.2a. It is remarkable that the entire evolution of the polarization response during cycling has manifested itself very distinctly in the third harmonic whose phase angle $\delta_3$ during certain transition time made a turn of about 180°. The result in Fig. 2b will be discussed in Section 4.2.

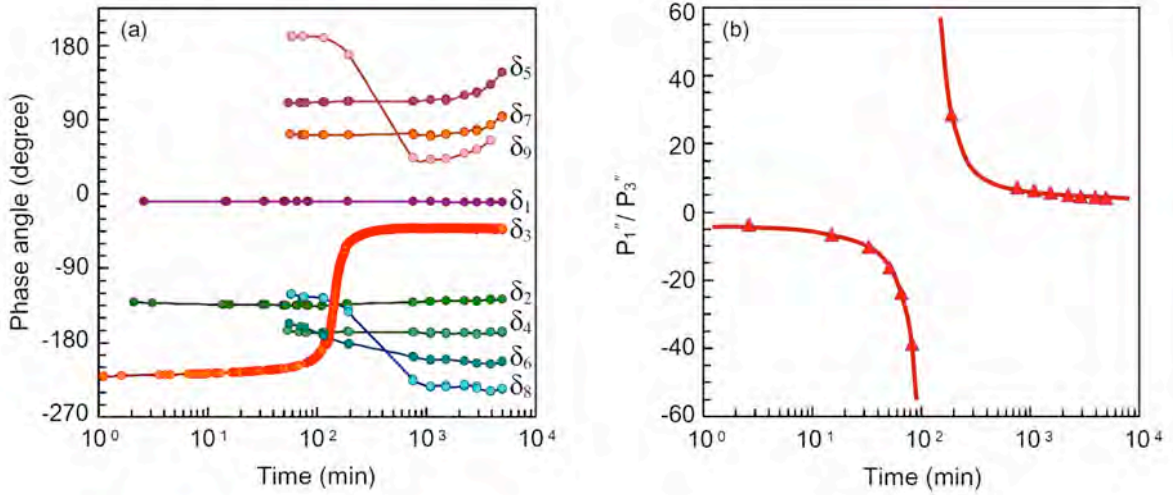

*Figure 2.* *Evolution of nonlinear parameters of polarization response of hard 58/42 PZT ceramics doped with 1.0 at.% Fe during cycling with electric field with amplitude of 25 kV/cm, at 1kHz (see also Fig 1a): (a) phase angles $\delta_n$ of the first nine harmonics and (b) out-of-phase amplitude ratio of the first and the third harmonic as a function of time.*

Behavior of $\delta_3$ presented in Fig. 2a can be understood using Fig.3, Eq.(1) and Table I. For a pinched hysteresis loop, the out-of-phase component of the third harmonic is positive and it contributes to the constriction of the loop at zero field. When the loop is relaxed, the out-of-phase component of the third harmonic changes sign and opens the loop at zero field (Fig. 3). This is exactly what is observed experimentally, as shown in Fig. 2a: during depinching $\delta_3$ changes in a step-like fashion from the second ($P_3' < 0, P_3'' > 0$) to the fourth ($P_3' > 0, P_3'' < 0$) quadrant of the complex $P_3' - P_3''$ plane. Interestingly, the phase angle of the first harmonic almost does not change during the field cycling (Fig.2,a). The small increase in its absolute value corresponds to the expansion of the whole hysteresis loop. The evolution of the third harmonic, on the other hand, clearly reveals the "depinching process": the loop evolves from the pinched loop (in the second quadrant), where the third harmonic suppresses the loop at zero field to the depinched loop (in the fourth quadrant) where third harmonic contributes to the expansion of the loop at zero field.

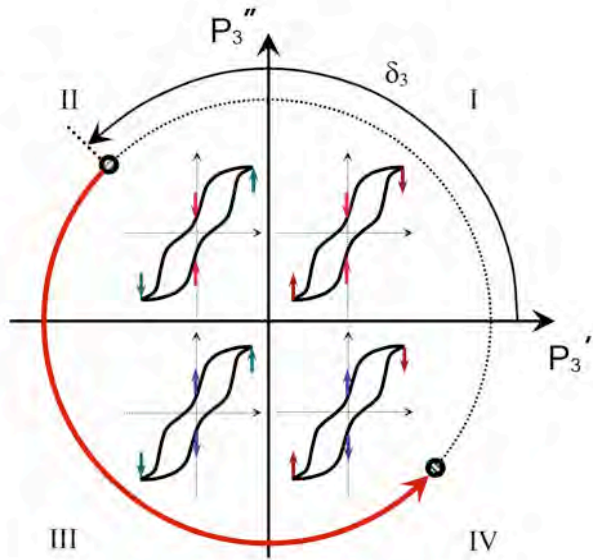

*Figure 3. Illustration of the third harmonic contribution to the total polarization response in different quadrants of the complex plane. Arrows indicate the suppressing (arrows down) or enhancing (arrows up) effects of the third harmonic on the hysteresis loop form*

The thermal activation nature of the deageing (or loop depinching) process by field cycling is demonstrated in Fig.4. The time $\tau$, which characterizes the transition from pinched to depinched loop, can be determined as the time when $\delta_3$ makes turn of about 180° (Fig.4a) or, which is equivalent, as the time corresponding to the point where the in-phase and out-of-phase components of $P_3$ change sign (Fig.4b).

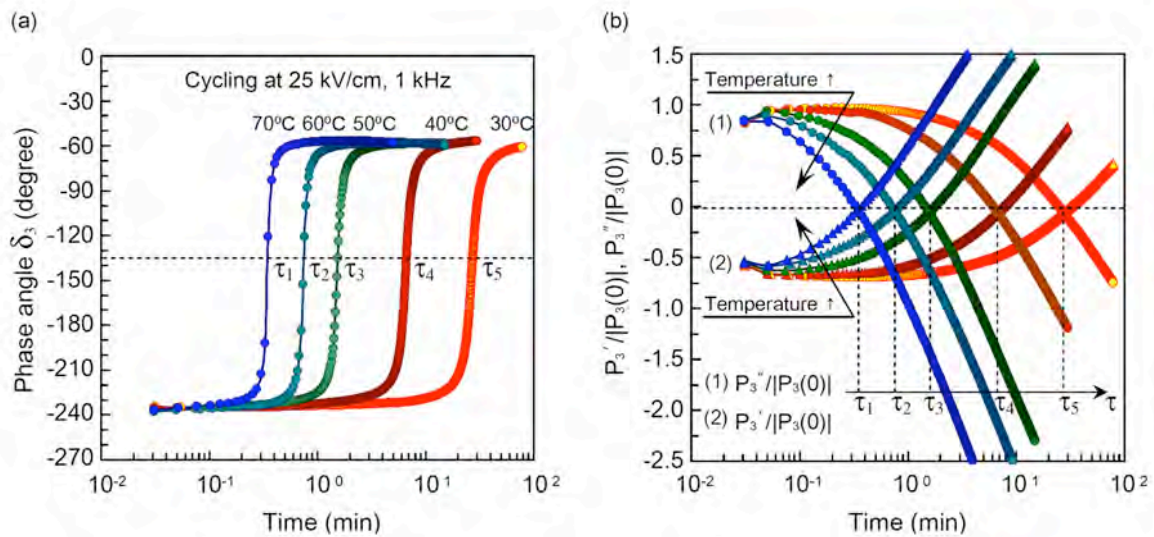

**Figure 4.** *Time dependence of $\delta_3$ of hard 58/42 PZT ceramics doped with 0.5 at%Fe under ac field cycling (25 kV/cm, 1 kHz) at various temperatures: (a) third harmonic phase angle as a function of time and temperature; (b) evolution of the in-phase and out-of-phase components of the third harmonic of polarization response with respect to their initial values measured at various temperatures.*

As seen from Fig 4a, the turning point of $\delta_3$ is at around -135° i.e., in the third quadrant when $P'_3 \approx P''_3 < 0$. Fig 4b shows that this happens when $P'_3 \approx P''_3 \approx 0$, i.e. when the third harmonic contributes very little to the nonlinearity of the loop. Clearly, at this point higher harmonics may still contribute to the loop shape but their effect appears to be small.

To ascertain that it is the time that characterizes the hysteresis relaxation process and not the number of cycles, the experiments have been performed at various frequencies. As shown in Fig. 5, the effect of frequency and, hence, number of cycles, on $\tau$ is relatively minor. On the other hand, as shown in Figs. 6 and 7, amplitude of the electric field and temperature have a strong effect on the relaxation time. As reported in Ref. 3, the relaxation time in Mn-doped PZT decreases exponentially with increasing field amplitude $[\tau \propto \exp(\alpha/E)]$. That observation has been made in the driving field range from 25 to 53 kV/cm. In this work, linear dependence of relaxation time has been observed in a hard (0.5 at% Fe) ceramics in the field range from 15 to 23 kV/cm (Fig.7).

There are several possible reasons for the different behavior of the relaxation time as a function of the driving field amplitude reported here and in Ref. 3, including the following. First, the field ranges are different (ours from 15 to 23 kV/cm, in Ref. 3 from 25 to 50 kV/cm.). A clear transition in the behavior of $\tau$ is seen in our measurements between 20 and 25 kV/cm, which could indicate a crossover from the linear to exponential regime. Second, the time in Ref. 3 is derived from the merging of the split switching current peaks and thus represents vanishing of the ageing effect(s) that contributed to the pinching at zero field. In our case, the relaxation times are derived from a parameter that characterizes different nonlinear processes during depinching. Note that $\tau$ characterizes the point in time at which nonlinearity is minimized ($|P_3| \approx 0$). It should be noted that the field ranges given here and in Ref. 3 are relative and may vary depending on ceramic composition and temperature or even in the way how ceramics are prepared. Moreover, it is very unlikely that linear dependence of relaxation time on field observed here holds down to zero field amplitude, since the finite relaxation time at zero field has no meaning; on the contrary, the ageing takes place.

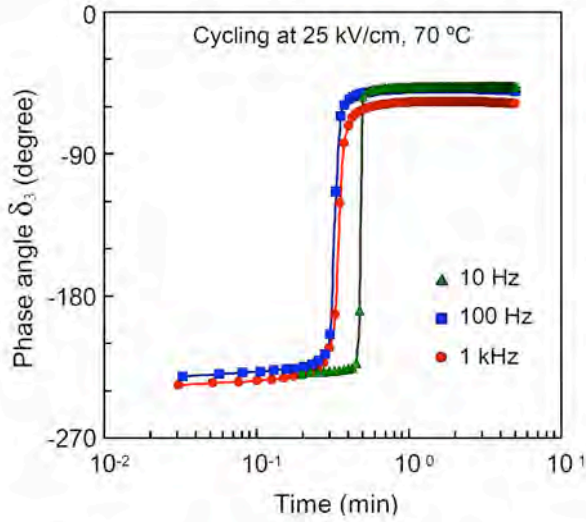

*Figure 5. Phase angle of the third harmonic of polarization response $\delta_3$ as a function of time for hard 58/42 PZT ceramics doped with 0.5at% Fe and cycled by electric field of 25 kV/cm at 72 °C for various frequencies.*

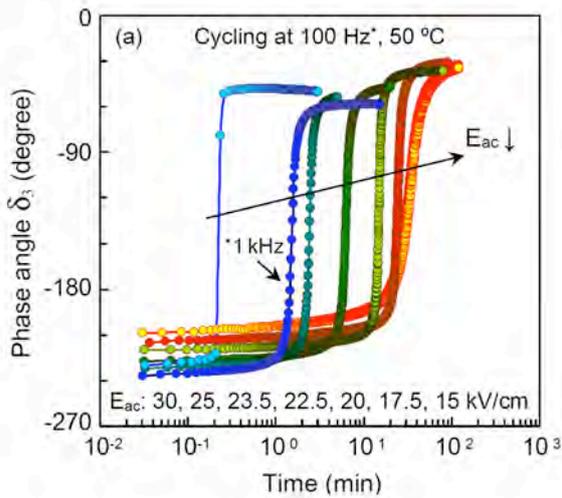 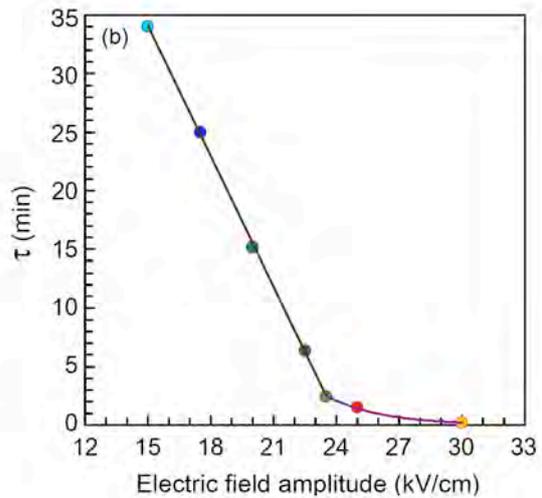

*Figure 6. (a) Evolution of the $\delta_3$ as a function of time for hard 58/42 PZT ceramics doped with 0.5 at% Fe during cycling at 50 °C by ac electric fields of various amplitudes. Note that the field amplitude decreases from left to right. (b) Dependence of the transition time $\tau$ on amplitude of the electric field.*

In the temperature range investigated the relaxation time is well described by the Arrhenius behavior (Fig. 7c). Importantly, the activation energy $E_{act}$ determined by the temperature dependence of $\tau$ plotted in Arrhenius scales and assuming $\tau \propto \exp(E_{act}/kT)$ relation, is found to be field dependent (Fig.7). For 58/42 PZT sample doped with 0.1 at% Fe the activation energy is 0.64 eV at 10 kV/cm, increasing to 0.92 eV at 15 kV/cm. The former energy corresponds well to the range of activation energies (0.56 to 0.7 eV) measured in Mn and Fe doped PZT by Carl and Härdtl.

The increase of the activation energy at higher fields could hint to involvement of long range charge migration in the loop relaxation;[35] this conjecture needs to be examined further. Our measurements of *dc*- and *ac*-conductivity in hard PZT[8] indicate activation energy of around 1 eV for direct current and 0.6 to 0.9 eV for alternating current conductivity. Since the depinching process involves migration of charged defects responsible for stabilization of domain structure (e.g., reorientation of dipoles and/or migration of charges near interfaces) one can propose the following picture.

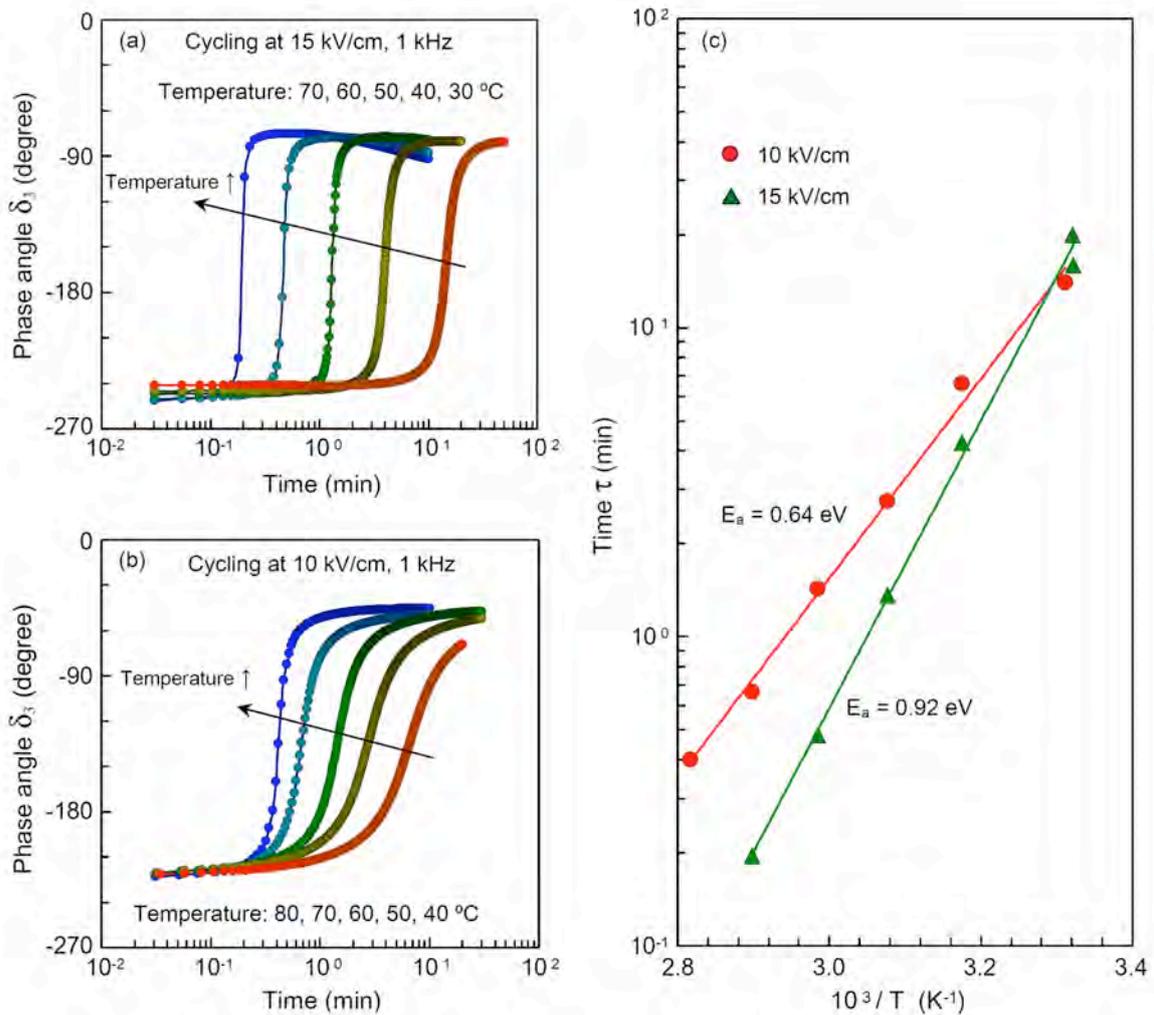

**Figure 7.** *Phase angle of the third harmonic of polarization response as a function of time for hard (0.1 at.% Fe-doped) PZT (58/542) ceramics cycled at various temperatures at ac-electric field amplitudes of (a) 15 kV/cm and (b) 10 kV/cm. (c) Temperature dependences of the corresponding transition times plotted in Arrhenius scales.*

As mentioned in Introduction, among different mechanisms that can contribute to hysteresis pinching (and depinching) are (i) reorientation of micro-dipoles eg., $Fe_{Ti} - V_O$ dipoles (so-called volume effect)[3,9,11,13] and (ii) the charge drift mechanism where macro-dipoles are created by longer-range migration of charges near interfaces (so-called grain boundary or surface effect).[3,10,19] The activation energy for each mechanism is clearly different, the first mechanism involving hopping of charges (e.g. oxygen vacancies or other charges associated with acceptor doping) over small distances, on the order of the lattice constant. The second mechanism involves displacement of charges over longer distances (fraction of the grain size or near-electrode layer). The hopping is associated with *ac*-conductivity and lower activation energy whereas migration of charges at interfaces may be related to the *dc*-conductivity and higher activation energy. The difference in activation energies for different materials, and the variation in the temperature and field dependence of relaxation times mentioned above can be then interpreted in terms of competing effects between two mechanisms contributing to the loop depinching. Assuming such a scenario, evolution of the third harmonic with time, temperature and field can be analyzed in more detail. The temporal evolution of $\delta_3$ in Figs. 6 and 7 shows that the transition between pinched to depinched state is faster (the step from -240° to -60° is sharper) at higher fields and higher temperatures. The reason for this could be that at higher fields and temperatures both mechanisms of depinching are active because available energy is sufficiently high to rearrange both micro-dipoles through short range charge hopping and "macro-dipoles" through large scale migration of charges; consequently the complete deageing happens faster.

Finally, it is interesting that frequency dependence of the $\delta_3$ transition (Fig. 5) also supports supposition that both longer-range and short range rearrangements of charges contributes to the deageing: the transition is faster (probably more complete) at the lower frequency (10 Hz) where more than one mechanisms may be active than at higher frequencies (100 Hz and 1 kHz) where shorter range charge movement dominates.

### 4.2. Characterization at subswitching fields

In the previous section evolution of the phase angle of the third harmonic of the polarization response $\delta_3$ was examined during cycling samples with fields near or above macroscopic switching fields. Such treatment leads to deageing of aged samples (macroscopically manifested by loop depinching) and this process appears to be associated with rearrangement of charges within samples. In this Section we consider $\delta_3$ as a function of the driving field amplitude in aged samples and in samples with thermally induced charge disorder. Besides the obvious extension of the work discussed in the previous section, additional motivation for this study is the following.

It is known that in soft PZT the dielectric hysteresis at weak and moderate fields can be well described using so-called Rayleigh law.[21,30,36,37] This law is valid for systems in which interfaces, such as domain walls, move in a spatially random energy landscape. Clearly, the Rayleigh law cannot be applied to well aged or poled hard materials were domain structure is configured by alignment (or ordering) of micro- or macro-dipoles with polarization within domains. The very notion of hysteresis pinching at zero field is incompatible with the Rayleigh law whose essence is hysteresis opening at zero field. Non-applicability of the Rayleigh law to hard materials has been demonstrated experimentally[29,30] and description of the pinched hysteresis within the Preisach framework was proposed in Ref. 38. However, when an aged hard sample is deaged the defects are randomized and the sample should show some characteristics of the Rayleigh behavior. In this Section we investigate if such transition from non-Rayleigh to Rayleigh-type behavior indeed happens in hard materials.

In the case of an ideal Rayleigh system, the Fourier expansion of the polarization response is given by:[39]

$$P = \chi(E_0)E_0 \sin(\omega t) + \sum_{n=1,3,5,}^{\infty} \frac{4\alpha_\chi E_0^2 \sin(n\pi/2)}{\pi n(n^2-4)} \cos(n\pi\omega t) \qquad (2)$$

where $\chi(E_0) = \chi_{init} + \alpha_\chi E_0$ is the dielectric susceptibility of the nonlinear ferroelectric, $\chi_{init}$ is the initial (zero field) susceptibility, and $\alpha_\chi$ is the Rayleigh coefficient. This representation of the Rayleigh relations has a number of features useful for analysis of the polarization response. In particular, it implies that in the ideal case the phase angle of the first harmonic $|\delta_1|$ must be lower than 23°, that phase angles of all higher odd-numbered harmonics are ±90°, and that ratios of out-of-phase amplitudes of any odd-numbered harmonics are field independent (e.g. $P_1''/P_3''= 5$). The absence of in-phase components in higher harmonics in (2) means that the nonlinear response is purely hysteretic (i.e. irreversible) and thus Eq.(2) clearly cannot describe the reversible response (or loop pinching) typical for aged hard materials at weak fields. As shown below, this may change if hard material is disordered by thermal or electric field treatment.

The dependences of the phase angles of the first and third harmonics on the subswitching driving field amplitude are shown in Fig. 8 for hard PZT ceramics in aged and quenched states. In aged ceramics the defect dipoles (micro and/or macro) conform with the polarization direction within domains leading to strong restoring force at zero field during loop cycling.[9] Therefore, $\delta_1$ is small and a relatively weak field applied for a short time during measurements cannot relax the hysteresis loop. In quenched ceramics, disorder of micro- and macro-dipoles at temperatures above $T_C$ is assumed to be frozen during the fast cooling to room temperature. The ageing sets on immediately upon cooling; however, if measurements are made well before the ageing process is completed the dipoles in the sample may be considered as at least partly randomized and should exhibit some characteristics of the Rayleigh behavior. Indeed, as seen from Fig. 8a, in quenched samples $|\delta_1|$ increases with increasing field, as would be expected in a system with randomly distributed pinning centers;[22] the increase is much weaker in the aged sample. The difference in behavior of $\delta_1$ of the quenched and the aged sample is, however, only quantitative. The major contribution to the polarization response is in-phase; the increase of the $|\delta_1|$ with increasing field amplitude corresponds to the increase of the Rayleigh contribution to what is usually called dielectric loss. Note that in the quenched sample $\delta_1$ approaches -23°, the maximum value expected from the Rayleigh relation (2).

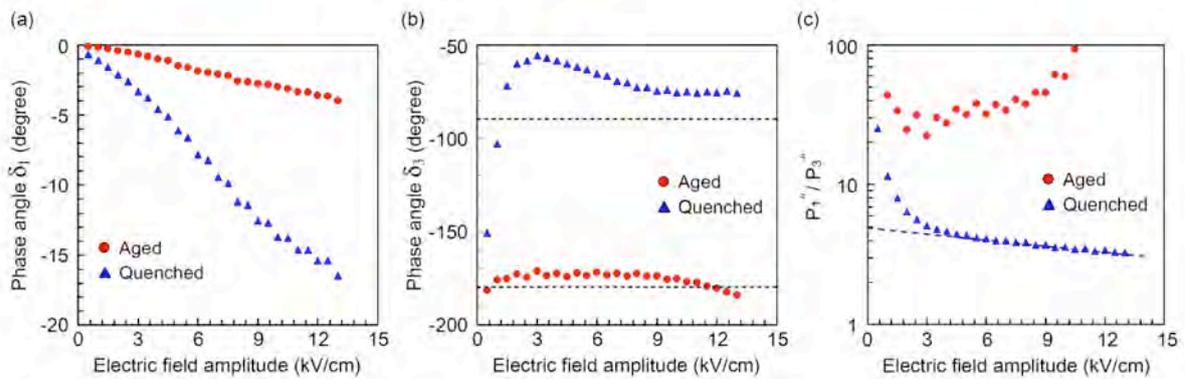

*Figure 8. Phase angles of the (a) first and (b) third harmonics of polarization response, and (c) ratios of the out-of-phase amplitudes of the first and third harmonic of polarization response as functions of electric field amplitude (1 kHz) for aged and quenched hard 58/42 PZT ceramics doped with 1.0 at.% Fe. The dashed lines indicate in (a) -90° and -180° and in (c) extrapolated amplitude ratio of 5 at zero field.*

In contrast to $\delta_1$, the observed difference in $\delta_3$ of aged and quenched materials is qualitative and more revealing (Fig. 8b): $\delta_3$ changed after quenching from mainly in-phase[b] (in aged material) to mainly out-of-phase (in quenched sample). This change in $\delta_3$ is consistent with the dominantly hysteretic response in disordered materials, as expected from Eq. (2). In addition, the ratio of out-of-phase components of the first and third harmonics ($P_1''/P_3''$) in quenched samples is approaching 5 as required by Eq. 2 (Fig. 8c; compare with Fig. 2b at long times), Together, these two results hint to transformation of the ferroelectric response of hard materials towards Rayleigh (soft) behavior during deageing by quenching. Interestingly, some properties of Eq. (2) are also observed in samples relaxed by switching fields where in relaxed state $P_1''/P_3'' \to 5$ (see Fig. 2b). However, it is unlikely that all features of the Rayleigh relation (2) hold at macroscopic switching fields where domain structure is completely reconstructed during cycling.[27]

---
[b] In terms of hysteresis behavior, the phase lag by either 0° or 180° corresponds to anhysteretic response.

While it has been shown above that aged hard samples show qualitatively different nonlinear response from quenched ones, it is interesting to see whether application of relatively weak subswitching fields may also frustrate ordering of charges in well aged samples and provoke some, even if only temporary, tendency towards disordering. We next show that by observing the third harmonic evolution during cycling with subswitching fields it is possible to detect such small changes toward disorder in aged samples. How much disorder can be induced at subswitching fields clearly depends on many parameters, including the degree of hardness, cycling time, field strength and temperature.

In case of slightly hard ceramic (doped with 0.1% Fe), some changes in behavior occur already during the first cycle (Fig.9a). The phase angle $\delta_3$ reaches the vicinity of -90° with increasing driving field amplitude during the first cycle and remains in this region (except near zero field) with decreasing field amplitude at all subsequent cycles. For a harder sample (with 0.5at% Fe), Fig.9.b, $\delta_3$ does not reach the region of -90° either with increasing field amplitude (where $\delta_3$ is nearly independent on field amplitude) or with decreasing field; however, $\delta_3$ noticeably evolves towards -90° during the descending branch of the cycle and at each run the sample demonstrates a sort of "memory" with respect to the state reached in the previous cycle. We propose that the phase angle (i.e. hysteresis) is higher during the decreasing than during the increasing branch of each cycle because sample gets locally softer (disordered) during increasing field and domain walls get trapped in newly created energy minima when the field is decreased. In other words, the motion of domain walls in the locally pre-disordered environment becomes partly irreversible and thus more hysteretic.

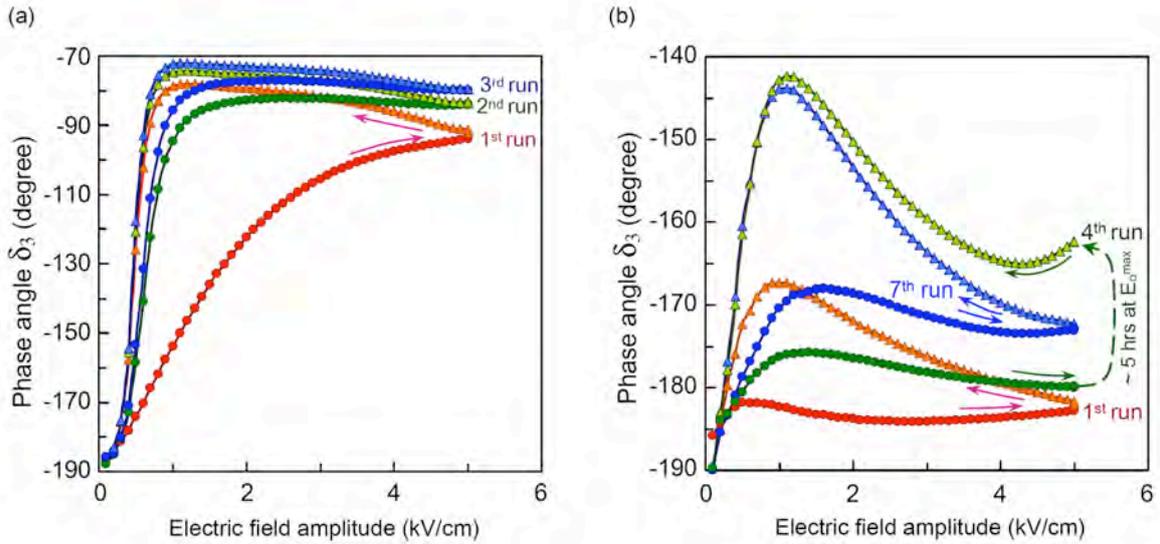

*Figure 9. Phase angle of the third harmonic of polarization response as a function of electric field amplitude (1 kHz) for hard 58/42 PZT ceramics doped with (a) 0.1% Fe and (b) 0.5% Fe. Arrows indicate the branches corresponding to increasing and decreasing amplitude of the driving field. For clarity only the 1$^{st}$, 4$^{th}$ and 7$^{th}$ cycle are shown. During the fourth cycle the sample was held 5 hours at maximal field amplitude at the end of the increasing branch. Once the field was decreased and the next cycle started, some aging occurred.*

If one accepts the bulk scenario for hardening, the driving force for ageing is interaction of dipoles (electric and elastic) with polarization and crystal lattice deformation and ensuing energy minimization.[9,15,40,41] The opposite is true for deageing: energy must be supplied to break interaction of polarization and lattice strain and dipoles; this process involves domain structure reconfiguration and charge re-distribution. At subswitching fields deaging (disordering) probably happens only within the short distance where domain walls travel during the cycling. This assumes that electric field may reorient only those dipoles that are within the region of reversed polarization. The complete deaging is thus not expected at subswitching fields at room temperature. However, if the subswitching field is applied at higher temperatures the additional energy is available. If the thermal energy, assisted with that supplied by the subswitching field, is sufficient to break the dipole-polarization interaction, total deaging of the sample at subswtching fields may be possible.

## 5. Summary

Ageing-deageing transition in acceptor doped PZT ceramics has been investigated by analyzing evolution of the phase angle of the third harmonic of the polarization response, $\delta_3$, with time, temperature and driving field amplitude. The correlations between $\delta_3$ and the hysteresis loop geometry (e.g., pinching) are analyzed in detail.

The experimental observations clearly demonstrate the link between the evolution of $\delta_3$ and the hysteresis loop transformation from pinched to depinched state during the deageing processes. The change of $\delta_3$ by ~90° at subswitching and ~180° at switching conditions corresponds to the transition between the aged (harden) and deaged (soften) states of the Fe-doped PZT. A detailed analysis of deageing during electric field cycling under switching conditions suggests presence of two or more mechanisms for domain wall stabilization, with activation energies ranging from ~0.6 to ~1.0 eV. It is proposed that the mechanism corresponding to lower activation energies is consistent with short-range charge hopping while higher activation energies could indicate longer-range migration of charges; the charge transport process possibly has common features with ionic conduction. At subswitching fields and in thermally quenched samples, the evolution of $\delta_3$ with driving field indicates behavior typical for soft materials, which can be qualitatively well described by the Rayleigh formalism. This is not the case for aged samples where Rayleigh behavior is not expected to be valid and is not observed experimentally. However, cycling at relatively small field amplitudes reveals *tendency* of aged samples to deage partially. This process reveals itself macroscopically in rotation of $\delta_3$ towards the values characteristic for soft or entirely relaxed hard materials.

**Acknowledgments**

The authors acknowledge financial support from the Swiss National Science Foundation and helpful discussions with Doru Lupascu.